\documentstyle[preprint,eqsecnum,aps]{revtex}

\begin{document}
\draft
\title{Covariant Variable Front Quark Model}
\author{Adam Szczepaniak, Chueng-Ryong Ji and Stephen R. Cotanch}
\address{Department of Physics, North Carolina State University, Raleigh, North Carolina 27695-8202}
\date{\today}
\maketitle
\begin{abstract}
We study the implications of Lorentz symmetry for
hadronic structure by formulating a manifestly covariant constituent quark
model and find
full covariance for any Lorentz transformation
requires utilizing a variable quantization surface.
Even though time-reversal invariance mandates a unique orientation
for the surface's direction,
we investigated alternative choices and find
small numerical sensitivity to different directions.
We also calculate
the static properties of the low lying pseudoscalar mesons
and compare results with existing noncovariant calculations. We conclude
that there is a little sensitivity of the calculated observables
 to the choice of the formalism, provided relativity 
 is properly implemented.
\end{abstract}
\narrowtext

\section{INTRODUCTION}
\label{sec:Intro.}

The explicit quark picture of
hadrons has been investigated by an increasing
 number
of models using effective, constituent quarks as the
fundamental
degrees of freedom~\cite{quarkmodels,Dzi,Coester}.
Despite the fact that a formal
link is missing between such QCD inspired models and exact QCD
 there are many indications that the constituent quark representation
does 
emerge as a result of strong vacuum correlations similar to
those leading to chiral symmetry breaking~\cite{Shifman}.
 Central to these approaches is
  a globally gauge
 invariant potential confining the quarks.
Justification for an
 effective potential is provided by the flux tube configurations
generated by the gluon field. To be realistic, these models
 must also incorporate relativity~\cite{Bogolubow,van} due to
 the large velocities associated with quarks having masses of
the order of few hundred MeV but bound on a typical hadronic scale of the
order of a $\mbox{GeV}$.
 The problem of formulating relativistic dynamics for a
fixed number of particles originated with the pioneering work
of
Dirac~\cite{Dirac} and has been extensively studied over the years on both
classical and quantum levels. Classically this is a Cauchy problem to
determine the world lines in Minkowski space, while in the quantum case one
seeks the probability amplitude distributions, as an initial value problem
on a 3-dimensional spacelike or lightlike surface. We refer the reader to
the extensive  literature on the subject~\cite{Smirnow,others,saz}. The
possible choices of the initial surface are commonly classified according to
the dimension of the corresponding stability group consisting
of purely kinematical transformations within the subspace defined by the
initial conditions. Conveniently, the group representation does not require
a full solution of the evolution equations. With this classification scheme
the surface of the light cone, also referred to as the light front, is very
appealing because it generates the largest stability group. At the field
theoretic level there are also practical reasons for quantizing on the light
front~\cite{Winberg}. At short distances color interactions can be treated
perturbatively and such separations, which are lightlike in a Minkowski
metric, usually dominate high energy processes~\cite{DIS}. Further, the QCD
vacuum is rigorously decoupled from excited Fock states when formulated on
the light
cone and this permits a more sensible quark Fock state expansion for hadron
states~\cite{LCFT}. However, until a formal connection is established
between field theory and QCD inspired models, such arguments should
cautiously be regarded.

A major deficiency of all formalism
choosing a fixed quantization surface, including the light
cone, is the lack of complete Lorentz covariance. Formalisms using a
specific
quantization surface defined by a fixed vector $n^\mu$ introduce an incorrect
functional dependence for all physical observables, e.g. form
factors,
upon $n^\mu$. A covariant wave function that describes a system of $N$
particles depends on $N$ independent {\it time} variables, $(n\cdot x)_i,
i=1\cdots N$, which in turn requires specifying the full dynamics by
prescribing the action of $N$ Hamiltonians on the initial wave function
spanning a $3\times N$ dimensional surface $\Sigma$~\cite{saz},
\begin{equation}
 \Sigma \equiv \{ \{x^\mu_1\cdots x^\mu_N\} : (n\cdot x_1)=
\cdots (n\cdot x_N) = 0\}\,. \label{sigma}
\end{equation}
If full covariance is important or desired, the dynamical evolution of a
system must be independent from the choice of
 quantization surface.
As a consequence, there is
no preferred
quantization scheme nor can one classify different choices by the
size of the corresponding stability group. Of course full covariance
is
not always necessary, provided strong
arguments exist for disposing of
unphysical degrees of freedom or spurious form factors arising in a
noncovariant formulation~\cite{LCFT,Karmanov}. In general, however, it
will be 
 important to identify the relevant
 degrees of freedom in a framework independent way.
The main objective of this paper is to compare relativistic calculations of
observables
for pseudoscalar mesons, for which the constituent quark structure is well
established, using both  noncovariant and covariant schemes and to document
framework
sensitivity.

In the following section the basic assumptions underling the
quark model calculation of physical observables are reviewed.
 Section~III addresses two
 common noncovariant schemes, the instant and the
light cone, while
 Sec.~IV details a new covariant quark model
based on a momentum dependent, variable front quantization.
Finally, numerical results are presented in Sec.~V
with major findings summarized in Sec.~VI.

\section{Mesonic properties in a constituent quark model}

All meson observables investigated in this work are specified
 in terms of hadronic matrix elements of local operators, having
 one of two forms

\begin{eqnarray}
{\cal A}_1 & \equiv &
\langle 0| {\overline \psi}(0)\Gamma \psi(0) | P \rangle, \nonumber \\
{\cal A}_2 & \equiv & \langle P' | {\overline \psi}(0)\Gamma \psi(0)
| P \rangle.  \label{mel}
\end{eqnarray}
Here $|P \rangle$ denotes a pseudoscalar meson state, having
4-momentum $P^\mu$, 
the fermion field operators ${\overline \psi}, \psi$
 refer to
the QCD quark fields and $|0\rangle$ denotes the vacuum. In the constituent
quark picture a meson state is
entirely represented by a valence constituent $q{\overline q}$ pair and the
exact QCD field operators are
 replaced by the effective
constituent fields~\cite{mock}.
 We shall comment on
these approximations in
Section~VI.

Since {\it time} parameters do not explicitly appear in the matrix
elements in Eq.~(\ref{mel}), only
knowledge of the single-time wave functions are needed.
A flat spacelike surface is defined by
a timelike 4-vector $n^\mu$,

\begin{equation}
n^\mu = (n^0, {\bf n})\;, n^2 = 1.
\end{equation}
For each particle $i$ described by
 configuration
 or momentum space 4-vector $A^\mu$ we define an
associated transverse 4-vector $A_T$ in the $i$-th
 subspace
of $\Sigma$ and longitudinal component $A_L$ by the projections,
$A^\mu = (A_L, {\bf A_T})$,

\begin{equation}
 A_L \equiv A\cdot n\;,
 A^\mu_T \equiv A^\mu - A_L 
n^\mu\ = (0,{\bf A}_T)\;. \label{spacelike}
\end{equation}
Since $A_T\cdot n = 0$, there are only three nonvanishing
components of $A_T$ denoted in 3-vector form by
$A_T = {\bf A}_T$.  The scalar product is

\begin{equation}
A\cdot B = A_L B_L + A_T \cdot B_T = A_L B_L + {\bf A}_T\cdot {\bf B}_T.
\end{equation}
Thus if $x$, and $p$ are conjugate,

\begin{equation}
[x^\mu, p^\nu] = ig^{\mu\nu},
\end{equation}
then the 3-dimensional transverse components,  $x_T$ and $p_T$, are
also canonically conjugate.
When $\Sigma$ is a lightlike surface
a null vector $n,n^2= 0$, must be introduced,

\begin{equation}
n^\mu = (1,{\bf n})\;, {\bf
n}^2= 1,
\end{equation}
with transverse and longitudinal variables

\begin{equation}
x_T = (x^- \equiv x^0 - {\bf n}\cdot {\bf x}, {\bf
x}_\perp)\;,
x_L = x^+ \equiv  x^0 + {\bf n}\cdot{\bf x},
\end{equation}
\begin{equation}
p_T = (p^+ \equiv p^0 + {\bf n}\cdot {\bf p}, {\bf
p}_\perp)\;,
p_L = p^- \equiv  p^0 - {\bf n}\cdot{\bf p},
\end{equation}
and 2-dimensional perpendicular components

\begin{equation}
{\bf A}_\perp \equiv {\bf A} - ({\bf n}\cdot {\bf A}){\bf n}. 
\end{equation}
for $A$ being either $x$ or $p$.

Returning to the fermion field operators $\psi$ and ${\overline \psi}$,
these can be canonically
quantized  on the surface $\Sigma$ and represented by
 a Hilbert space expansion in terms
of
operators $b(k_T,\lambda,\tau), d(k_T,\lambda,\tau)$; for $n^2 > 0$

%\begin{eqnarray}
%& & \psi(x_L=0,x_T)  =  \int {{d^3k_T}\over {(2\pi)^3}}
%{m\over {k_L}} \sum_{\tau,\lambda=\pm} \nonumber \\
%& & \left[ b(k_T,\lambda,\tau)u(k_T,\lambda)e^{-ik_Tx_T}
% +  d^{\dag}(k_T,\lambda,\tau)v(k_T,\lambda)e^{i
%k_Tx_T} \right] \nonumber \\
%& & {\overline \psi}(x_L=0,x_T)  =  \int {{d^3k_T}\over
%{(2\pi)^3}} {m\over {k_L}} \sum_{\tau,\lambda=\pm} \nonumber \\
%& & \left[ b^{\dag}(k_T,\lambda,\tau){\overline u}(k_T,\lambda)e^{i
%k_Tx_T}
% + d(k_T,\lambda,\tau){\overline v}(k_T,\lambda)e^{-i
%k_Tx_T} \right], \nonumber \\ \label{exp}
%\end{eqnarray}

\widetext

\begin{eqnarray}
& & \psi(x_L=0,x_T)  =  \int {{d^3k_T}\over {(2\pi)^3}}
{m\over {k_L}} \sum_{\tau,\lambda=\pm}
\left[ b(k_T,\lambda,\tau)u(k_T,\lambda)e^{-ik_Tx_T}
 +  d^{\dag}(k_T,\lambda,\tau)v(k_T,\lambda)e^{i
k_Tx_T} \right], \nonumber \\
& & {\overline \psi}(x_L=0,x_T)  =  \int {{d^3k_T}\over
{(2\pi)^3}} {m\over {k_L}} \sum_{\tau,\lambda=\pm}
\left[ b^{\dag}(k_T,\lambda,\tau){\overline u}(k_T,\lambda)e^{i
k_Tx_T}
 + d(k_T,\lambda,\tau){\overline v}(k_T,\lambda)e^{-i
k_Tx_T} \right]. \nonumber \\ \label{exp}
\end{eqnarray}
\narrowtext
$\lambda$ is a helicity component, $\tau$ represents other quark 
quantum numbers, i.e. flavor and color,
$k_L = \sqrt{m^2 + |k^2_T|}$, $m$ being the constituent quark mass and
$u,v$ Dirac spinors
in a helicity basis. The normalization conditions for the spinors,
creation and annihilation operators are summarized in the Appendix.
The quark, antiquark ($q{\overline q}$) wave function $\Psi$ is defined as
the probability
amplitude for finding the $q{\overline q}$ pair with given quantum numbers
within the meson state,

\begin{eqnarray}
& & (2\pi)^3\delta^3(P_T - \sum_{i}^2 k_{i T})
\Psi^\alpha(k_{1T},k_{2T};\lambda_1,\lambda_2;\tau_1,\tau_2)
= \nonumber \\
& & \sqrt{ {m_1\over {k_{1 L}}}}\sqrt{  {m_2\over {k_{2 L}}}} {1\over
{\sqrt{2P_L}}} \langle 0| d(k_{2 T},
\lambda_2,\tau_2)b(k_{1 T},\lambda_1,\tau_1)| P_T;\alpha \rangle\;
\nonumber \\
\label{wf}
\end{eqnarray}
where

\begin{eqnarray}
& & |P_T;\alpha> = \sum_{\lambda,\tau}\int [dk_{iT}]^{q{\overline q}}_P
\sqrt{ {m_1\over {k_{1 L}}}}\sqrt{  {m_2\over {k_{2 L}}}} \nonumber \\
 & & \sqrt{2P_L}\Psi^\alpha(k_{iT};\lambda_i;\tau_i)
 b^{\dag}(k_{1T},\lambda_1,\tau_1)
 d^{\dag}(k_{2T},\lambda_2,\tau_2)| 0\rangle,   \nonumber \\
& & [dk_{iT}]^{q{\overline q}}_P =  [dk_{iT}]_P \equiv \prod_{i}^2{{d^3k_{iT}}\over {(2\pi)^3}}
 (2\pi)^3 \delta^3(P_T - \sum_{i}^2 k_{iT}).\nonumber  \\
\label{state}
\end{eqnarray}
Here $P_T$ is the transverse component of the meson momentum,
$P_L = \sqrt{M_\alpha^2 + |P^2_T|}$ and $M_\alpha$ being the
pseudoscalar meson
mass  ($\alpha=1\dots 8$ specify the low-lying
octet state).
The states are normalized according to

\begin{equation}
\langle P';\alpha|P;\beta \rangle = \delta_{\alpha\beta} 2(2\pi)^3 P_L
\delta^3(P'_T - P_T)
\end{equation}

\begin{equation}
\sum_{\lambda,\tau} \int [dk_{iT}]_P \Psi^{\dag \alpha}(k_{iT};\lambda;\tau)
\Psi^{\dag \beta}(k_{iT};\lambda;\tau) = \delta_{\alpha\beta}. \label{norm}
\end{equation}
Combining Eqs.~(\ref{mel}),~(\ref{exp}) and ~(\ref{wf}) yields

\widetext

\begin{eqnarray}
{\cal A}_1 & = & \langle 0|{\overline \psi}(0)\Gamma \psi(0)
|P_T;\alpha\rangle \nonumber \\
& = & \sum_{\lambda,\tau}\int[dk_{iT}]_P
\sqrt{2P_L}\Psi^\alpha(k_{iT};\lambda_i;\tau_i)
 \left [
\sqrt{ {m_1\over {k_{1 L}}}} {\overline v}(k_{2T},\lambda_2)
\Gamma_{\tau_1,\tau_2}
\sqrt{  {m_2\over {k_{2 L}}}} u(k_{1T},\lambda_1)
\right], \label{a1}
\end{eqnarray}

\begin{eqnarray}
{\cal A}_2 & = &\langle P';\alpha | {\overline \psi}(0)\Gamma\psi (0)
| P;\beta \rangle\nonumber \\
& = & \sum_{\lambda\lambda'\tau\tau'}\int
[dk_{iT}]_{P'}[dq_{iT}]_P
\sqrt{2P'_L} \Psi^{\dag \alpha}(k_{iT};\lambda'_i;\tau'_i)
 \sqrt{2P_L} \Psi^\beta(q_{iT};\lambda_i;\tau_i)
\nonumber \\
& &
\left[  \left[
\sqrt{ {m_1\over {k_{1 L}}}} {\overline u}(k_{1T},\lambda'_1)
\Gamma_{\tau'_1,\tau_1}
\sqrt{  {m_1\over {q_{1 L}}}} u(q_{1T},\lambda_1)
\right]
\delta_{\tau'_2\tau_2}\delta_{\lambda'_2\lambda_2}(2\pi)^3\delta^3(k_{2T}-q_{2T})
\right. \nonumber \\
& - & \left.
 \left[
\sqrt{ {m_1\over {k_{2 L}}}} {\overline v}(k_{2T},\lambda'_2)
\Gamma_{\tau'_2,\tau_2}
\sqrt{  {m_2\over {q_{2 L}}}} v(q_{2T},\lambda_2)
\right]
\delta_{\tau'_1\tau_1}\delta_{\lambda'_1\lambda_1}(2\pi)^3\delta^3(k_{1T}-q_{1T})
\right]. \label{a2}
\end{eqnarray}
\narrowtext
The expressions for the null plane quantization, $n^2=0$, can similarly be
derived and only differ due to 
 different normalization conditions for the light cone states. 
To obtain the null plane expressions the factors $1/\sqrt{k_{iL}}$
should be replaced by $1/\sqrt{k_i^+}$.

\section{Noncovariant Constituent Quark Models}

In this section we examine the matrix elements ${\cal A}_i$ for two
 noncovariant schemes corresponding to different
choices of the quantization surface $\Sigma$.
Even though the matrix elements are formally 
 Lorentz covariant functions of external hadron
momenta, when modeled by wave
functions
defined on a given surface, i.e. for fixed $n^\mu$, they in general  fail
to maintain this property. Covariance of the
matrix element, say ${\cal A}_2$, evaluated for an operator $\Gamma$ 
that contains $k$-Lorentz indices 
means that for a general Lorentz transformation
$\Lambda$ 

\begin{eqnarray}
& & {\cal A}^{\mu_1\cdots\mu_k}_2(P',P)  =
 \langle P'| {\overline \psi}(0)\Gamma^{\mu_1\cdots\mu_k}\psi (0)
 | P \rangle \nonumber \\ &  & = \Lambda^{\mu_1}_{\nu_1}\cdots
\Lambda^{\mu_k}_{\nu_k}
 {\cal A}^{\nu_1\cdots\nu_k}_2(\Lambda^{-1}P',\Lambda^{-1}P).
 \label{cov}
 \end{eqnarray}
If $U(\Lambda)$ denotes the representation of the transformation $\Lambda$
in the space of physical states $|P\rangle $ then Eq.~(\ref{cov})
follows from the transformation property of the quark fields
 under $U(\Lambda)$

\begin{equation}
U(\Lambda)\psi(0)U^{-1}(\Lambda) = S(\Lambda)\psi(0),
\end{equation}
where $S(\Lambda)$ belongs to the matrix representation of the
Lorentz group in the space of bispinors. In general the
representation of
$U(\Lambda)$ in the Fock space of constituents 
 involves interactions which have non-zero matrix elements
between Fock states with different particle number. Thus any
truncation of the Fock space to a finite
 number of particles violates covariance and the evaluation
of the matrix elements ${\cal A}_i$ 
 becomes frame dependent. For this reasons existing
approaches~\cite{quarkmodels,LCFT} adopt specific frames,
 depending on the particular quantization surface
$\Sigma$ used.
Obviously, the nonrelativistic situation  is much simpler
because matrix element
 in a truncated Fock space  
remain covariant under galilean
symmetry.

\subsection{Instant quantization}

Instant quantization is defined by a fixed  timelike vector
$n^\mu, n^2=1$. Without loss of generality we orient the
coordinate system such that 

\begin{equation}
n^\mu = (1,\vec{0})
\end{equation}
where the arrow symbolizes the three transverse
components of a given four vector in the instant quantization.
A standard assumption in the instant constituent quark model is that
the ground state pseudoscalar meson wave function in the meson rest frame, 
defined by $\hat{P} \equiv (M,\vec{0})$, 
has $S=L=0$~\cite{PDG} and can be written as~\cite{quarkmodels}

%\begin{eqnarray}
%& & \Psi^a(\vec{k}_i;\lambda_i;\tau_i) = \nonumber \\
%& & i[{\lambda^a\over \sqrt{2}} \otimes {I\over \sqrt{3}}]_{\tau_1\tau_2}
%{1\over {\sqrt{2}}}\left[\delta_{\lambda_1+}\delta_{\lambda_2-}
%- \delta_{\lambda_1-}\delta_{\lambda_2+}\right]\Phi(\vec{q}^2/\beta^2(m_i))
%\end{eqnarray}

\begin{eqnarray}
& & \hat{\Psi}^a_{in}(\vec{\hat{k}}_i;\hat{\lambda}_i;\tau_i) =
\chi^\alpha_{\tau_1\tau_2}
\hat{\xi}^{\hat{P}}(\hat{\lambda}_i),
\hat{\Phi}_{in}(\vec{\hat{k}}_i), \nonumber \\
& & \chi^\alpha_{\tau_1\tau_2} \equiv i\left[{{\lambda^a}
\over {\sqrt{2}}}\otimes {{I}\over \sqrt{3}}\right]_{\tau_1\tau_2}
\nonumber \\
& & \hat{\xi}^{\hat{P}}(\hat{\lambda}_i) \equiv {1 \over
\sqrt{2}}\left[\delta_{\hat{\lambda}_1+}\delta_{\hat{\lambda}_2-}
- \delta_{\hat{\lambda}_1-}\delta_{\hat{\lambda}_2+}\right].\label{instcm}
\end{eqnarray}
Here $\lambda^a$ denote the Gell-Mann matrices in the $SU(3)$ flavor
space,
 $I$ is the identity in the color space and
$\hat{\Phi}_{in}(\vec{\hat{k}}_i)$
is the instant orbital wave function usually assumed gaussian

\begin{equation}
\hat{\Phi}_{in}(\vec{\hat{k}}_i) =
N\exp{\left[- { { (\eta\vec{\hat{k_1}} + (1-\eta)\vec{\hat{k_2}})^2 }\over
{2\beta_{in}^2}}\right] },\label{orbin}
\end{equation}
with $\beta^2_{in}\sim\sqrt{m_1m_2/(m_1+m_2)}\beta^2$, $\beta$
being the flavour independent harmonic oscillator size parameter,
$\eta = m_1/(m_1+m_2)$ and $N$ determined by Eq.~(\ref{norm}),

\begin{equation}
\int [d\vec{\hat{k}}_i]_{\hat{P}}|\hat{\Phi}_{in}(\vec{\hat{k}}_i)|^2 = 1.
\end{equation}
The caret symbol denotes a quantity evaluated in the
meson rest frame, $\vec{\hat{k}}_1 + \vec{\hat{k}}_2 = \vec{\hat{P}} =
0$.

From Eqs.~(\ref{state}) and ~(\ref{cov}) it now follows that Lorentz boosts,
not rotations, are no longer covariant in the truncated instant wave function
basis.
Thus the 
 the matrix
element ${\cal A}_1$ involving the wave function $\hat{\Psi}^a_{in}$
should only be evaluated in the center of momentum frame. However, to
calculate ${\cal A}_2$
 wave functions are needed for
both
initial and final mesons having  
momenta $P^\mu$ and $P'^\mu$ respectively.
In principle interaction dependent boost operators provide the
connection between these wave functions, however,
without knowing the underlying relativistic dynamics in the
valence constituent Fock space sector alternative connections are necessary.
 The standard assumption
is that the relation between wave functions describing states with
different 4-momenta  can be approximated by the one following from
the noninteracting case.
Within such an approximation
the rest frame spin component $\hat{\xi}^{\hat{P}}(\hat{\lambda}_i)$
of the wave function
transforms under an active boost to a frame in which the meson has
momenta
 $\vec{0} \to \vec{P} = \vec{k}_1 + \vec{k}_2$  according
to~\cite{Wigner}

\begin{eqnarray}
& & \hat{\xi}^{\hat{P}}(\hat{\lambda}_i) \to
\xi_{in}^{P}(\vec{k}_i;\lambda_i)
= \sum_{\hat{\lambda}} D^{1/2}_{\lambda_1\hat{\lambda}_1}(\vec{k}_i;1)
D^{1/2}_{\lambda_2\hat{\lambda}_2}(\vec{k}_i;2)
\hat{\xi}^{\hat{P}}(\hat{\lambda}_i), \nonumber \\
\label{inst}
\end{eqnarray}
where $D^{1/2}(\vec{k}_i;i) = D^{1/2}(R_W(\Lambda(\hat{P} \to P));i)$ is the
spinor matrix representation of the Wigner rotation $R_W$ acting 
on the i-th particle spinor,

\widetext

\begin{equation}
D^{1/2}_{\lambda_1\lambda_2}(\vec{k}_i;i) =
{ { (E_i(\vec{k}_i) + m_i)(\hat{E}_i(\vec{k}_i) + m_i) +
\vec{k}_i\cdot \vec{\hat{k}}_i +
i\sigma\cdot(\vec{k}_i\times\vec{\hat{k}}_i) }
\over {\sqrt{2(E_i(\vec{k}_i) + m_i)(\hat{E}_i(\vec{k}_i)+m_i)(m_i^2 +
E_i(\vec{k}_i)\hat{E}_i(\vec{k}_i) + \vec{k}_i\cdot\vec{\hat{k}}_i) } } }, \label{wig}
\end{equation}

%\narrowtext

and

\begin{eqnarray}
& & E_i(\vec{k}) \equiv \sqrt{m_i^2 + \vec{k}^2}\;,
{\cal M}(\vec{k}_1,\vec{k}_2) \equiv \sqrt{ (E_1(\vec{k}_1) + E_2(\vec{k}_2))^2
- \vec{P}^2}, \nonumber \\
& & \hat{E}_i(\vec{k}_i) = E_i(\vec{k}_i) {{E_1(\vec{k}_1) +
E_2(\vec{k}_2)}\over { {\cal
M}(\vec{k}_i) } }  - { \vec{P} \over {{\cal M}(\vec{k}_i)}}\cdot \vec{k}_i,
\nonumber \\
& & \vec{\hat{k}}_i
 = \vec{k}_i - {{ \vec{k}_i\cdot\vec{P}
}\over
\vec{P}^2 }\vec{P}   + \vec{P} {{E_1(\vec{k}_1) + E_2(\vec{k}_2)}\over
{ {\cal M}(\vec{k}_i) } }\left( \vec{k}_i\cdot {\vec{P} \over \vec{P}^2
}\right) -   \vec{P} {{E_i(\vec{k}_i)}\over  {{\cal
M}(\vec{k}_i)}}.
\end{eqnarray}
\narrowtext
Note that orbital wave function $\Phi_{in}(\vec{k}_i)$ for the meson
state with 3-momentum $\vec{P}$,
 is obtained from the rest frame wave function
 $\hat{\Phi}_{in}(\vec{\hat{k}}_i)$
by replacing $\vec{\hat{k}}_i \to \vec{k}$.
Inserting the rest frame wave function
$\hat{\xi}^{\hat{P}}(\lambda_i)$ of 
Eq.~(\ref{instcm}) in Eq.~(\ref{inst}) and summing over helicities
$\hat{\lambda}$
yields

%\begin{eqnarray}
%& & \xi_{in}^{P}(\lambda_i)  =  \sqrt{2}{\sqrt{m_1m_2}\over {\sqrt{{\cal
%M}^2(\vec{k}_i) - (m_1-m_2)^2}}}\nonumber \\
%& \times & {\overline u}(\vec{k}_1,\lambda_1)\gamma_5
%v(\vec{k}_2,\lambda_2).\label{xitot}
%\end{eqnarray}

\begin{equation}
\xi_{in}^{P}(\vec{k}_i;\lambda_i)  =  \sqrt{2}{\sqrt{m_1m_2}\over
{\sqrt{{\cal M}^2(\vec{k}_i) - (m_1-m_2)^2}}}
 {\overline u}(\vec{k}_1,\lambda_1)\gamma_5
v(\vec{k}_2,\lambda_2).\label{xitot}
\end{equation}
The general structure of the above equation, i.e. the appearance of the
$u$ and $v$ Dirac spinors, is due to the proportionality of
 $\xi^{P}_{in}(\vec{k}_i;\lambda_i)$
to
 the probability amplitude of finding a
free ${q\overline q}$ pair with 4-momenta $k_i$ and helicities $\lambda_i$
in the state with zero spin and 4-momentum $k_1 + k_2$,

\begin{equation}
\xi^{P}_{in}(\vec{k}_i;\lambda_i) \propto \langle \vec{k}_i;\lambda_i |
\vec{k_1} + \vec{k_2} \rangle.
\end{equation}
Thus $(k_i\cdot
\gamma \pm m_i)\xi^{P}_{in}(k_i;\lambda) = 0$ leading to the form of
 Eq.~(\ref{xitot}).

\subsection{Light cone quantization}

In a light cone formulation one can choose a coordinate system
such
that the null vector $n^\mu,n^2=0$, specifying the light cone surface
$\Sigma$ of Eq.~(\ref{sigma}),
has coordinates

\begin{equation}
n^\mu = (1,0,0,1).
\end{equation}
Since the surface $\Sigma$ is invariant under any boost
  the 
 wave function can be obtained in a particular Lorentz frame
 from the wave function
in any other frame by a pure Lorentz
boost using only kinematical
transformations~\cite{Smirnow}.
However, rotations around the $x$ and $y$ axis, unlike the instant case will
 now depend upon the dynamics. Hence matrix elements calculated for
non-zero
spin hadrons will not respect the fundamental relations for different spin
projections that follow from rigorous covariance~\cite{Karmanov,Deuteron}.

To obtain a model light cone wave function however, i.e. without solving  
dynamical equations of the underlying theory, some approximations are
necessary. In an exact approach
hadron states
can be expand in either  an instant or a light cone quantized basis, i.e.

\begin{eqnarray}
& & |\vec{P}\rangle_{in} = \sum_N \sum_{\lambda\tau}
\int [d\vec{k}_i]^N_P \Psi^N_{in}(\vec{k}_i;\lambda_i;\tau_i)
| N,\vec{k}_i;\lambda_i;\tau_i \rangle_{in} \nonumber \\
& & |\tilde{P}\rangle_{lc} = \sum_N \sum_{\lambda\tau}
\int [d\tilde{k}_i]^N_P \Psi^N_{lc}(\tilde{k}_i;\lambda_i;\tau_i)|N,
\tilde{k}_i;\lambda_i;\tau_i \rangle_{lc}. \label{comp}
\end{eqnarray}
Here the tilde symbolizes transverse momentum components
 in the light cone quantization and subscripts $in$ and $lc$ refer 
 to instant and light cone basis respectively.
If the relation between instant
$|N,\vec{k};\lambda_i
\rangle_{in}$ and light cone, $|N, \tilde{k}_i;\lambda_i \rangle_{lc}$
basis states is known, the light cone wave functions $\Psi^N_{lc}$
can be obtained from the instant ones, $\Psi^N_{in}$ using
Eq.~(\ref{comp}).
This relation, however, cannot be established on the basis of kinematical
considerations alone, i.e. without knowledge of a full dynamics.
In the noninteracting case, for example, the light cone
basis is obtained from the instant one by separately boosting
each constituent having momenta $\vec{k}_i$ in the instant state to its
respective rest frame, then performing a boost to the
infinite
momentum frame followed by a boost back to the original frame of the
particle. 
Since boosts in the instant basis will in general involve
interactions, the above mentioned approximations must be made
in implementing these boosts~\cite{Coester,Smirnow}.

The light cone wave function is usually obtained under
 an interaction free approximation for the boost transformations
similar to the procedure for instant wave
functions
in different Lorentz frames as discussed in the previous section. Then the
 spin component $\xi_{lc}^{P}(\tilde{k}_i;\lambda_i)$ of
the complete constituent $q{\overline q}$ light
cone wave function

\begin{equation}
\Psi^{\alpha}_{lc}(\tilde{k};\lambda_i;\tau_i)  = 
\chi^\alpha_{\tau_i}\xi_{lc}^P(\tilde{k}_i;\lambda_i)\Phi_{lc}(\tilde{k}_i)
\end{equation}
is obtained from the rest frame spinor wave function
$\xi^{\hat{P}}(\lambda_i)$ of Eq.~(\ref{instcm}) by,
($P^+ = k^+_1 + k^+_2, {\bf P}_\perp = {\bf k}_{\perp 1} + {\bf k}_{\perp
2}$)~\cite{ruscy,Coester,Dzi},

\begin{equation}
\xi_{lc}^P(\tilde{k}_i;\lambda_i) = \sum_{\hat{\lambda}}
 U^{1/2}_{\lambda_1\hat{\lambda}_1}(\tilde{k}_1)
U^{1/2}_{\lambda_2\hat{\lambda}_2}(\tilde{k}_2)
\xi^{\hat{P}}(\hat{\lambda}_i) ,\label{bla}
\end{equation}
where $U^{1/2}(\tilde{k}_i)$ are the Melosh rotations representing the
product of the three boosts described above in the $i$-th
particle spinor space,

\begin{equation}
U^{1/2}(\tilde{k}_i) = { { m_i + {{k^+_i}\over {P^+}}{\cal M}(\tilde{k}_i)
+ i\epsilon_{rs}\sigma_r \left( {\bf k}_{\perp s i} - {{k^+_i}\over {P^+}}
{\bf P}_{\perp s} \right) }
\over
{ \sqrt{(m_i +  {{k^+_i}\over {P^+}}{\cal M}(\tilde{k}_i))^2 +
 ({\bf k}_\perp - {{k^+_i}\over {P^+}} {\bf P}_\perp)^2 } } } 
 \end{equation}
and

\begin{eqnarray}
& & {\cal M}(\tilde{k}_i) = \sqrt{ P^+(k^-_1(\tilde{k}_1) + k^-_2(\tilde{k}_2))
- {\bf P}_\perp^2 } \nonumber \\
& & =   P^+ \left[ {{m_1^2 + k^2_{1\perp}}\over
{k^+_1}} + {{m_2^2 + k^2_{2\perp}}\over {k^+_2}} \right].
\end{eqnarray}
This leads to the light cone spinor wave function of the form

%\begin{eqnarray}
%& & \xi_{lc}^{P}(\lambda_i)  =  \sqrt{2}{\sqrt{m_1m_2}\over {\sqrt{{\cal
%M}^2(\tilde{k}_i) - (m_1-m_2)^2}}}\nonumber \\
%& \times & {\overline u}(\tilde{k}_1,\lambda_1)\gamma_5
%v(\tilde{k}_2,\lambda_2) \label{spinlc}
%\end{eqnarray}

\begin{equation}
 \xi_{lc}^{P}(\tilde{k}_i;\lambda_i)  =
\sqrt{2}{\sqrt{m_1m_2}\over {\sqrt{{\cal M}^2(\tilde{k}_i) - (m_1-m_2)^2}}}
{\overline u}(\tilde{k}_1,\lambda_1)\gamma_5
v(\tilde{k}_2,\lambda_2) \label{spinlc}
\end{equation}
which, not surprisingly, is identical to the instant one of
Eq.~(\ref{xitot})
with the Dirac spinors expressed in terms of the light cone momenta
since they have been obtained under the same dynamical approximations.
 One can easily show that in the free case

\begin{equation}
\xi_{lc}^{P}(\tilde{k}_i;\lambda_i) = \sum_{\lambda'}
  U^{1/2}_{\lambda_1\lambda'_1}(\tilde{k}_1)
U^{1/2}_{\lambda_2\lambda'_2}(\tilde{k}_2)
\xi_{in}^{P}(\vec{k}_i;\lambda'_i) \label{ala}
\end{equation}
and that

\begin{equation}
U^{1/2}_{\lambda\lambda'}(\tilde{k}) = {\overline
u}(\tilde{k};\lambda)u(\vec{k};\lambda').\label{ola}
\end{equation}
Eq.~(\ref{spinlc}) immediately follows form
Eq.~(\ref{ala}),(\ref{ola}) and Eq.~(\ref{xitot}).

Finally, the orbital light cone  wave function
$\Phi_{lc}(\tilde{k}_i)$ is 
treated independently  from the instant one
($\Phi_{in}(\vec{k}_i)$) and is usually assumed to be a
 a function of the free invariant $q{\overline q}$ mass ${\cal
M}(\tilde{k}_i)$, 

%\begin{eqnarray}
%& & \Phi_{lc}(\tilde{k}_i) =
%\Phi_{lc}({\cal M}^2(\tilde{k}_i)) =  \nonumber \\
%& & \Phi_{lc}\left( P^+ \left[ {{m_1^2 + k^2_{1\perp}}\over {k^+_1}} 
%+ {{m_2^2 + k^2_{2\perp}}\over {k^+_2}} \right] \right). \label{lcwf}
%\end{eqnarray}

\begin{equation}
                                                            
\Phi_{lc}(\tilde{k}_i) =
\Phi_{lc}({\cal M}^2(\tilde{k}_i)) =
\Phi_{lc}\left( P^+ \left[ {{m_1^2 + k^2_{1\perp}}\over {k^+_1}} 
+ {{m_2^2 + k^2_{2\perp}}\over {k^+_2}} \right] \right). \label{lcwf}
\end{equation}

\section{Covariant Constituent Quark Model}

Even though the valence wave functions describing states with different
momenta have been related by a Lorentz transformation, the hadronic
matrix elements still do not have the
 proper transformation properties.  
This is due to the interaction free boost approximation in the truncated
Fock space and leads to an unphysical $n^\mu$ dependence for the matrix
elements.
 In a fixed quantization surface scheme the components of quantization
vector $n^\mu$ remain fixed as one changes the reference frame
i.e. performs a Lorentz which change the values of the components of
particle momenta. As the hadronic matrix elements in general
depend on products like $n \cdot P_i$, where $P_i$ denotes any
particle momenta, the change of the reference frame will lead to
 different values for the matrix elements. As an example
consider the matrix element ${\cal A}_1$. In a noncovariant 
 fixed-$n^\mu$ model the pion matrix element with the pseudovector
current $\Gamma^\mu$ has an incorrect form

\begin{equation}
{\cal A}^\mu_1 = \langle 0|\Gamma^\mu |P\rangle =
f_1(P\cdot n)P^\mu + (n\cdot P) f_2(P\cdot n)n^\mu
\end{equation}
instead of 

\begin{equation}
{\cal A}^\mu_1 = f_\pi P^\mu. \label{cova1}
\end{equation}
In Fig.~1 we plot $f_1/f_\pi$ and $f_2/f_\pi$ with $f_\pi
=93\mbox{MeV}$ as functions
of the pion energy calculated in the instant (two solid lines)
and light cone (two dashed lines) schemes respectively.
The two upper and lower curves correspond to $f_1$ and $f_2$ respectively.
We refer to Sec. V for the details of the calculation of matrix elements.
It is seen that both functions are energy dependent in contrast to
the physical case of Eq.~(\ref{cova1}), and that $f_2 \ne 0$ over a wide
 region of the pion energy. It is also worth nothing
that as $E_\pi \to \infty$ the instant and light cone description
become identical which follows from their equivalence
in the infinite
 momentum frame.
Fig.~1 shows a large discrepancy in the predictions
for  a hadronic matrix element using different components of the current
 due to the luck of manifest covariance of a fixed-$n^\mu$ scheme.
 A very natural way to restore covariance for
 models with fixed
  number of constituents is to assume that the components of the
 4-vector $n^\mu$
are not fixed but rather are variable and transform covariantly under a
Lorentz
 transformation
 just like all other physical Lorentz
tensors leaving the scalar products $n\cdot P_i$ invariant under change of
the reference frame.  This can only by achieved if the vector $n^\mu$ is
parameterized  in terms
of  the physical tensors germane to the system~\cite{saz,ASAW}.
  Thus we introduce
 the concept of a
 variable front to maintain manifest covariance
in a constituent quark model calculation of  physical matrix elements, as
depicted in Fig~2.
With our prescription the front is variable
with 
 $n^\mu\propto P^\mu$ since $P^\mu$ is the only
4-vector available for ${\cal A}_1$. This immediately yields
 Eq.~(\ref{cova1})  for the decay constant without
further approximations.
Similarly, to derive an expression for
${\cal A}_2$ $n^\mu$ can be represented by a linear combination of
the intial and final  meson momenta,
$n^\mu\propto a P^\mu + b P'^\mu$  where $a$ and $b$ are in general
functions of $P\cdot P'$. The requirement of the time-reversal
invariance, however, further constrains $n^\mu$ and demands $a = b$.
We shall return to this point later and for now assume the
general case $a\ne b$. It is important to note that the cases
 $n^2 > 0$ and $n^2=0$ must be analyzed separately since each leads to
very different transverse variables
with attending integration measures.
Generally, non-zero mass particles lead to
$n^2 > 0$ for which all transverse variables are of the
form given by Eq.~(\ref{spacelike}).
We restrict our investigation to this case.

As shown in Section~III the 
$q{\overline q}$ wave function $\Psi_{cov}(k_{iT};\lambda_i)$,
depends on the individual quark helicity components through the
 representation

\begin{equation}
\Psi_{cov}(k_{iT};\lambda_i) \propto \sum_{\alpha\beta}{\overline
u}(k_{1T},\lambda_1)_\alpha\Gamma_{\alpha\beta}
v(k_{2T},\lambda_2)_\beta,
\end{equation}
where $\Gamma_{\alpha\beta}$ is a scalar constructed from Dirac
matrices, quark
momenta and $n^\mu$. As in Section~III we adopt the interaction
free approximation for the Lorentz spin rotations.
Then the full variable front $q{\overline q}$ wave function is
(see
 Eqs.~(\ref{xitot}) and ~(\ref{spinlc}))

\begin{eqnarray}
& & \Psi^\alpha_{cov}(k_{iT};\lambda_i;\tau_i)
  = \sqrt{2}\chi^\alpha_{\tau_i} {{ \sqrt{m_1m_2} }
 \over { \sqrt{{\cal M}^2(k_{i T}) - (m_1-m_2)^2} } } \nonumber \\
& \times & {\overline u}(k_{1T},\lambda_1)\gamma_5v(k_{2T},\lambda)
\Phi_{cov}(k_{iT}), \label{covspin}
\end{eqnarray}
with, ($k_{1 T} + k_{2 T} = P_T$),

\begin{eqnarray}
& & {\cal M}(k_{i T}) \equiv \sqrt{(k_{1 L} + k_{2 L})^2 - |P^2_T|},
\nonumber \\
& & k_{iL}(k_{iT}) \equiv \sqrt{m_i^2 + | k^2_{iT} |}. \label{invcov}
\end{eqnarray}
%Although the wave function $\Psi_{cov}(k_{iT},\lambda_i)$ does not
%transform 
% covariantly with respect to spinor indices, the matrix elements, ${\cal
%A}_i$  which involve summation over $\lambda_i$'s are manifestly
%covariant.
The spin independent component of the wave function,
$\Phi_{cov}(k_{i T})$ in Eq.~(\ref{covspin})
is a Lorentz invariant function of the individual momenta $k_{iT}$ which we
choose to express in terms
 of the  variable ${\cal M}(k_{i T})$,

\begin{equation}
\Phi_{cov}(k_{i T}) = \Phi_{cov}({\cal M}^2(k_{1T},k_{2T})).
\end{equation}
Thus $\Phi_{cov}({\cal M}^2)$ is the probability amplitude distribution
for the invariant $q{\overline q}$ 
mass. In a diagrammatical phenomenology, $\Psi^\alpha_{cov}({\cal M})$
describes  an off shell,
$\delta^2 \equiv  {\cal M}^2 - M_\alpha^2$ ($M_\alpha$ is the hadron mass),
 form factor for a vertex that conserves the momentum components $k_T$. 
Here we model $\Phi_{cov}({\cal M}^2)$ assuming the
 form
\begin{equation}
\Phi_{cov}({\cal M}^2) = N\exp\left[ {-  { {{\cal M}^2(k_{iT})} \over {8
\beta^2_{cov}}}}\right]. \label{covorb}
\end{equation}
which is a straightforward generalization of the quark model orbital wave
functions
discussed in the previous sections. If $n^\mu$ is chosen as
in the
instant form then
$k_T \to \vec{k}$  and in the rest frame $\Phi_{cov}({\cal M}^2) =
\Phi_{in}(\vec{k}_i)$ with $\beta^2_{cov} \sim
\beta^2_{in}(m_1+m_2)^2/4m_1m_2$.
Similarly, if  $n^\mu$ specifies a fixed light cone front, then

%\begin{eqnarray}
%& & \Phi_{cov}({\cal M}) = \Phi_{lc}(\tilde{k}_i) \nonumber \\
%& = & \Phi_{lc}\left( P^+ \left[ {{m_1^2 + k^2_{1\perp}}\over {k^+_1}} 
%+ {{m_2^2 + k^2_{2\perp}}\over {k^+_2}} \right] \right). \nonumber
%\end{eqnarray}

\begin{equation}
\Phi_{cov}({\cal M}) = \Phi_{lc}(\tilde{k}_i)
 = \Phi_{lc}\left( P^+ \left[ {{m_1^2 + k^2_{1\perp}}\over {k^+_1}} 
+ {{m_2^2 + k^2_{2\perp}}\over {k^+_2}} \right] \right).
\end{equation}

An important point to mention is that the wave function, $\Psi_{cov}$
of Eq.~(\ref{covspin}) always enters the calculation of any matrix element
with the sum over helicities, i.e.
\begin{equation}
 \Psi^\alpha_{cov}(k_{iT};\lambda_i;\tau_i)  \to
 \Psi^\alpha_{cov}(k_{iT};\tau_i)_{\alpha\beta} =
\sum_{\lambda_1\lambda_2}
u_\alpha(k_{1T},\lambda_1)\Psi^\alpha_{cov}(k_{iT};\lambda_i;\tau_i){\overline
v}_\beta(k_{2T},\lambda_2),
\end{equation}
where $\alpha,\beta$ are the Dirac indices.
Then it can be easily shown that
$\Psi^\alpha_{cov}(k_{iT};\tau_i)_{\alpha\beta}=
 \Psi^\alpha_{cov}(k_{i};\tau_i)$ i.e. it is the quantization surface
independent. We have thus shown that out prescription of constructing
the relativistic wave function is quantization surface independent.
The choice of a particular quantization surface, either by prescribing
a vector with fixed components $n^\mu$ or with components covariant
under Lorentz
transformations, corresponds to a different ways of choosing the transverse,
kinematical momentum components to study the same wave function.

Finally, we again stress that for
calculations of 
the matrix element ${\cal A}_1$ we shall use $n^\mu = P^\mu/P^2$,
 ($P^2 = M^2$) while for ${\cal A}_2$,
 we use the most general normalized form

\begin{eqnarray}
& & n^\mu = {1\over { ({1\over 4} - \kappa^2)Q^2 + M^2}} \left[ ({1\over 2} -
\kappa)P^\mu + ({1\over 2} + \kappa) P'^\mu \right],\nonumber \\
\label{ncov}
\end{eqnarray}
where $Q^2 = - (P'-P)^2$, and
 ${-1/2 < \kappa < 1/2}$ with time-reversal
invariance requiring $\kappa = 0$.

\section{Numerical Results}

All key observables can be obtained 
 from the matrix elements  ${\cal A}_1$ and 
${\cal A}_2$. Of special interest are the meson decay constants
and form factors. For the decay constant,  
the operator $\Gamma$ in the ${\cal A}_1$ matrix element of  Eq.~(\ref{mel})
 takes the form

\begin{equation}
\Gamma \to \Gamma^{\mu,\alpha}_{5 \tau_1\tau_2} =
[\gamma^\mu\gamma_5]{1\over 2}[\lambda^\alpha
\otimes I]_{\tau_1\tau_2}.
\end{equation}
giving for the pseudoscalar octet mesons
$\alpha = 1\dots 8$, with masses $M_\alpha$  

\begin{equation}
f_\alpha P^\mu\delta^{\alpha\beta} = -i\left[{\cal A}_1\right]_5^{\mu, \alpha\beta}.
\end{equation}
Using Eqs.~(\ref{a1}),~(\ref{covspin}) and $n^\mu = P^\mu/M$, with
 $P_T = 0$, and $k_{1T} = - k_{2T} \equiv q_T$, $q_T = q - (q\cdot
P)/M^2$, then explicitly yields 

\begin{eqnarray}
& & f_\alpha = \sqrt{6}\int {{d^3q_T}\over {(2\pi)^3}} {1\over \sqrt{M_\alpha}} \Phi^\alpha({\cal M}^2 (q_T,-q_T)) 
\nonumber \\
& \times & {{ k_{1L}(q_T)m_2 + k_{2L}(q_T)m_1 } \over 
\sqrt{k_{1L}k_{2L}({\cal M}^2(q_T,-q_T) - (m_1 - m_2)^2)} }. \label{fpion}
\end{eqnarray}
In the constituent quark model with valence quarks only the hadronic state
corresponds to a loosely bound system of massive, $m_q \sim 300\mbox{MeV}$
quarks. This model does not agree with the chiral structure of the pion
as described by strongly bound $q{\overline q}$ yielding a small
 pion mass.
 In order to be consistent, in the constituent quark model
one thus should use a pion mass of the order of $600\mbox{MeV}$. The mock
meson
prescription referred to in Sec. II then relate the decay constant $f_\alpha$
given by Eq.~(\ref{fpion}) to the physical one. Similarly, using the same
prescription  one
can define the quark condensation parameter, $\langle{\overline q} q\rangle
= \langle 0|{\overline q}q|0 \rangle$, by~\cite{SVZ}
\begin{equation}
i\sqrt{2}{{\langle 0|{\overline q}q | 0 \rangle }\over {f_\pi}}
= \langle 0| {\overline \psi}(0)\Gamma_5 \psi(0) | P,\pi
\rangle. \label{con}
\end{equation}
with
\begin{equation}
\Gamma \to \Gamma_{5\tau_1\tau_2} =
[\gamma_5]{1\over 2}[(\lambda^1 - i\lambda^2)
\otimes I]_{\tau_1\tau_2}.
\end{equation}
In our covariant front form model 
 $\langle {\overline q}q
\rangle $  is then

\begin{equation}
\langle {\overline q}q \rangle = - f_\pi\sqrt{6}\int {{d^3q_T}\over
{(2\pi)^3}} \sqrt{M_\pi} \Phi({\cal M}^2(q_T,-q_T)). \label{qqmod}
\end{equation}
It is worth mentioning that despite the uncertainties in the pion mass in
the constituent quark model Eqs.~(\ref{fpion}) and ~(\ref{qqmod}) imply that
the condensate becomes $M_\pi$ independent.

For the electromagnetic form factor $F_\alpha(Q^2)$, we use the current
operator 
 
\begin{equation}
\Gamma \to \Gamma^\mu_{e.m. \tau_1\tau_2} =
[\gamma^\mu]{1\over 2}[(\lambda^3 + {{\lambda^8}\over {\sqrt{3}}})
\otimes I]_{\tau_1\tau_2}, \label{em}
\end{equation}
in the matrix element ${\cal A}_2$ giving 
\begin{equation}
F_\alpha(Q^2)(P^\mu + P'^\mu) = \left[{\cal A}_2\right]^{\mu ,\alpha\alpha}.
\end{equation}
with $Q^2 \equiv (P' - P)^2$.
In our variable front the form factor is now  
 a function of the parameter
$\kappa$ defining  the quantization surface in terms of the initial
and final pion momenta (see Eq.~(\ref{ncov})). For $\kappa\ne 0$, the
formalism does violate both time reversal and
 gauge invariance, thus we shall restrict our analysis to $\kappa=0$.
We shall comment on the  $\kappa\ne 0$ case later.
For $\kappa = 0$
 the form factor reduces to

\widetext
\begin{eqnarray}
& & F_\alpha(Q^2) = \int {{d^3k_T} \over {(2\pi)^3}} 
\Phi^{\dag\alpha}({\cal M}'^2)\Phi^\alpha({\cal M}^2) \nonumber \\
& & \left[e_1 {{ k'_{1L}({\cal M}^2 - (m_1-m_2)^2) + k_{1L}({\cal M}'^2 -
(m_1-m_2)  + k_{2L}( (k_{1L} - k'_{1L})^2 - Q^2) } 
\over {2\sqrt{k_{1L}k'_{1L}}\sqrt{{\cal M}^2 - (m_1-m_2)^2}
\sqrt{{\cal M}'^2-(m_1-m_2)^2} } }  + (1 \to 2) \right], \nonumber \\
\label{fff}
\end{eqnarray}
where

\begin{eqnarray}
& & {\cal M}  = {\cal M}(k_T + \eta P_T, - k_T + (1-\eta)P_T)\nonumber \\
& & {\cal M}' = {\cal M}((k_T + (1-\eta)q_T) + \eta P'_T,
-(k_T + (1-\eta) q_T) + (1-\eta)P'_T), \nonumber \\
& & k_{1L}=k_{1L}(k_T + \eta P_T)\,,
k'_{1L}=k_{1L}((k_T + (1-\eta)q_T) + \eta P'_T),\nonumber \\ 
& &  k_{2L}=k_{2L}(- k_T + (1-\eta)P_T) = 
k_{2L}(- (k_T + (1-\eta)q_T) + (1-\eta)P'_T),
\end{eqnarray}
\narrowtext
with ${\cal M}(k_{iT})$ and $k_{iL}(k_T)$ given by Eq.~(\ref{invcov}),
$\eta = m_1/(m_1+m_2)$ and $e_1$, $e_2$ being the quark and antiquark 
charges, respectively. 

The results obtained in the covariant front model are summarized in Table~1
and Figs.~3-5. The table gives the predictions of the model for
the static properties of the low-lying pseudoscalar mesons 
as a function of the flavour independent wave function size parameter $\beta$.
As Table~1 indicates the value $\beta\sim 250\mbox{MeV}$ permits
a reasonable description of all static properties for both $\pi$ and $K$. 
In contrast, for the instant model  of Section IIIA it is not possible to
simultaneously reproduce both $\pi$ and $K$ decay constants and the
charge 
 radius 
 with a single value of $\beta$. In the instant 
model the pion charge radius can be reproduced for quark mass
and $\beta$ both being in a narrow range around $\sim 250
\mbox{MeV}$.
For these values fitting the $K^+$ charge radius requires a strange
quark mass of about  $450\mbox{MeV}$, however, the $K^0$
charge radius is then too small ,   
 $\langle r^2_{K^0} \rangle \sim -0.03 - -0.02\mbox{fm}$ as compared to 
$\sim -0.05\mbox{fm}$ experimentally.
Similarly the pion decay constant is problematic in the instant model since it is predominantly 
independent of the quark mass and 
to correctly model requires a $\beta$ value which yields an
unphysically small charge radius.
For additional insight we also have compared our results with the previous
light cone analyses of Dziembowski {\it et al.} Ref.~\cite{Dzi}, and
Coester {\it et al.} Ref.~\cite{Coester}.
The major difference between the two models is the treatment of  
 the spinor wave function of Eq.~(\ref{bla}) for which
Ref.~\cite{Dzi}
 replaces the invariant quark mass ${\cal M}$ by a constant value. 
This difference is not significant for the description of static
properties but as discussed below, is for the form factors. 
For the two noncovariant light cone approaches there is a narrow range
of $\beta$ values which correctly reproduce the static properties,
especially when compared to the instant model, but not as narrow as
in the covariant model.
Figs.~3 to 5 show our predictions for the $\pi,K^+$ and $K^0$
electromagnetic form factors for spacelike photon momentum. 
For the  pion form factor we also show model predictions for  the
instant (dotted line), 
Dziembowski {\it et al.} (dashed line) and
Coester {\it et al.} (dot-dashed line).
Notice that only the covariant front model (solid line) discussed here and
the light cone approach of Coester {\it et al.}~\cite{last} 
provide a reasonable description at high $Q^2$. 
In the Breit frame with orientation 
such that
$ q = (0,0,0,\sqrt{Q^2})$, 
 $n^\mu$ of Eq.~(\ref{ncov}) is $n^\mu=(1,0_T)$ and the
covariant model form factor reduces to the instant one except that the
relative
wave function now depends on the invariant quark mass ${\cal M}$. 
Thus the instant model deficiency steams from the nonrelativistic
treatment of the spin-independent wave function component.
 In contrast, the form factor shortcomings in the 
  Dziembowski 
model,  can be directly attributed to the spin wave function approximations.

All of the covariant model predictions for the form factors plotted in
Figs.~3 to 5 assume $\kappa = 0$. Even though
$\kappa\ne 0$ is unphysical it is useful to study the sensitivity
 of model predictions for the form factors as a function of $\kappa$
provided the unphysical aspects are corrected.
 Since variation of $\kappa$ represents changing
the quantization surface such analysis permits assessing
 the importance of covariance in a relativistic model calculation.
For $\kappa \ne 0$ current conservation can easily be
restored by the electromagnetic current modification

\begin{equation}
\Gamma_{e.m.}^\mu \to \left[ g^\mu_\nu - {{q^\mu q_\nu}\over {q^2}} \right]
\Gamma_{e.m.}^\nu.
\end{equation}
while the effects from time reversal violation does not affect
 the form factor for spin-0 particles. It would affect
 form factors for spin$>$0, e.g. in the deuteron case, and we shall
report on this case in a subsequent publications.
Fig.~6 documents the sensitivity of the covariant model pion form factor
 to the parameter $\kappa$.
The two curves establish the maximum variation of the form factor  as a
function of $\kappa$
and correspond to the two extremes, $\kappa = 0$ (solid line) and $\kappa = 1/2$ (dashed line) 
 ($F(Q^2,\kappa) = F(Q^2,-\kappa)$).
Notice that in this $Q^2$ region  the form factor is extremely
insensitive to $\kappa$, i.e. the choice of the quantization surface.

\section{Summary and Conclusions}

We have developed a covariant approach for investigating hadronic structure 
involving hadronic matrix elements of
constituent quark operators. The model incorporates all relevant symmetries,
Lorentz and gauge by construction and permits assessing the 
sensitivity of computed observables to the choice in 
quantization scheme (parameter $\kappa$).
As such our method can be regarded as a covariant criteria since it can be implemented 
in any quark model to detail sensitivity to front orientation. A large sensitivity indicates 
the need for full covariance and perhaps a more sophisticated hadronic model. 
We have also calculated various mesonic properties in this model and
compared the results with previous but noncovariant schemes.
Key findings are that relativity is crucial but that relativistic approaches which properly treat Wigner spin rotations and appropriate ansatzs for the relative , spin-independent wave function 
(dependence not directly on $k_{iT}$ but
through ${\cal M}(k_{i T})$ instead),
can achieve equivalent phenomenological descriptions at least for the present quality of data. 
In particular we have found that a proper relativistic quark model with
a soft (gaussian) wave function can account for most of the
normalization of the
electromagnetic form factors at intermediate $Q^2$. This is in agreement
with many recent findings~\cite{pionff} and indicates that the region
where
the perturbative QCD becomes dominant may be shifted to momentum transfers
as large as $Q^2 > 10-20\mbox{MeV}^2$.
As more precise data becomes available it may be possible
to distinguish alternative formulations including 
additional degrees of freedom such as effective gluonic excitations and/or exotic
quark configurations, as in principle expected in QCD.
Our  approach can be easily extended to address these new degrees
of freedom
to again assess the importance of covariance in a model independent way.

\acknowledgments
AS. would like to acknowledge D. Robson, A.~G. Williams and J .Piekarewicz
for many useful discussions. Financial support from U.S. D.O.E. grants 
DE-FG05-88ER40461 and DE-FG05-90ER40589 is also acknowledged.

\section{Appendix} 

We use the convention of Ref.~\cite{itzu}

\begin{eqnarray}
& & {\overline u}(k_T,\lambda)u(k_T,\lambda') =
\delta_{\lambda\lambda'}\;
{\overline v}(k_T,\lambda)v(k_T,\lambda') = -\delta_{\lambda\lambda'}\,,
\nonumber \\
& & \sum_\lambda u(k_T,\lambda){\overline u}(k_T,\lambda) =
{{k\cdot \gamma + m}\over {2m}},\nonumber \\
& & \sum_\lambda v(k_T,\lambda){\overline v}(k_T,\lambda) =
{{k\cdot \gamma - m}\over {2m}}. \nonumber
\end{eqnarray}
Creation and annihilation operators and single particle states are
canonically normalized according to,

\begin{eqnarray}
\{b(k'_T,\lambda'),b^{\dag}(k_T,\lambda)\} & = & (2\pi)^3 {{k_L}\over m}
\delta^3(k'_T,k_T), \nonumber \\
\{d(k'_T,\lambda'),d^{\dag}(k_T,\lambda)\} & = & (2\pi)^3
{{k_L}\over m}
\delta^3(k'_T,k_T), \nonumber \\
\langle k'_T,\lambda'|k_T,\lambda \rangle &  = & (2\pi)^3
{{k_L} \over m} \delta^3(k'_T - k_T). \nonumber
\end{eqnarray}

\newpage

%\begin{figure}
%\caption{Quantization surfaces for calculating of ${\cal A}_1$.
% Dashed line represents the instant scheme, dotted line 
%the light cone  and solid line the covariant variable front formalism.}
%\end{figure}
\figure{FIG.1. Pion decay constant in noncovariant scheme. Solid line
represents the instant calculation, dashes line the light cone one. }

\figure{FIG.2. Quantization surfaces for calculating of ${\cal A}_1$.
 Doted line represents the instant scheme, dashed-dotted line
the light cone  and dashed line the covariant variable front
formalism.}
%\begin{figure}
%\caption{ Pion electromagnetic form factor for spacelike $q^2 = -Q^2$
%momentum
%transfer. Solid line is the prediction of the covariant variable front
%model 
%with $m_u=m_d = 250\mbox{MeV}, \beta = 250\mbox{MeV}$. Data is taken from
%Ref.[21].}
%\end{figure}

\figure{FIG.3a,b. Pion electromagnetic form factor for spacelike $q^2 =
-Q^2$ momentum
transfer. Solid line is the prediction of the covariant variable front model
with $m_u=m_d = 250\mbox{MeV}, \beta = 250\mbox{MeV}$. Data is taken from
Ref.[25].}

%\begin{figure}
%\caption{ $K^+$ electromagnetic form factor. ($m_s = 480\mbox{MeV}$). Data
%is taken from Ref.[23].}
%\end{figure}
\figure{FIG.4a,b. $K^+$ electromagnetic form factor. ($m_s =
480\mbox{MeV}$). Data is taken from Ref.[26].}

%\begin{figure}
%\caption{ $K^0$ electromagnetic form factor. \hfill}
%\end{figure}
\figure{FIG.5a,b. $K^0$ electromagnetic form factor. \hfill}

%\begin{figure}
%\caption{ Sensitivity of the pion electromagnetic form factor to
%choice in quantization surface
% ($m_u=m_d=250\mbox{MeV}, \beta = 240 \mbox{MeV}$).}
%\end{figure}
\figure{FIG.6a,b. Sensitivity of the pion electromagnetic form factor to
choice in quantization surface
 ($m_u=m_d=250\mbox{MeV}, \beta = 240 \mbox{MeV}$).}

\newpage
\widetext
\begin{table}
\caption{Summary of the numerical results for the covariant variable front
model, ($m_u = m_d = 250\mbox{MeV}, m_s = 480\mbox{MeV}$).}

\begin{tabular}{|r|c|c|c|c|c|c|}
\hline
$\beta$ MeV  & $f_\pi$ & $f_K$ & $<q{\overline q}>^{(1/3)}$ & $<r^2_\pi>$ &
$<r^2_{K^+}>$ & $<r^2_{K^0}>$ \\
\hline
.20  &  75.8 &  87.0 & -176.  &     0.55  &    0.48  &      -0.111  \\

.22  &  83.1 &  96.5 & -190.  &     0.47  &    0.41  &      -0.089  \\

.24  &  90.1 & 105.7 & -205.  &     0.41  &    0.35  &      -0.073  \\

.26  &  96.8 & 114.7 & -218.  &     0.36  &    0.31  &      -0.060  \\

.28  & 103.3 & 123.4 & -231.  &     0.32  &    0.27  &      -0.051  \\
\hline
exp. & 93.2 &  113. &   -250.  &    0.43 &     0.34   &
-0.054 \\      &  MeV    & $\pm$ 2. MeV  &  $\pm$ 25. MeV &   $\pm$ 0.02
fm$^2$ &$\pm$ 0.05 fm$^2$ & $\pm$0.026  fm$^2$\\
\hline
\end{tabular}
\end{table}
\narrowtext

% Instant f_\H

%\begin{eqnarray}
%& & f_H = \sqrt{6}\int {{d^3\vec{\hat{q}}}\over {(2\pi)^3}}{1\over
%\sqrt{M_H}}\Phi(\vec{\hat{q}}^2/\beta^2(m_i)) \nonumber \\
%& & { { \sqrt{(\hat{E}_1 + m_1)(\hat{E}_2 + m_2)} +
%\sqrt{(\hat{E}_1-m_1)(\hat{E}_2-m_2)} } \over {2\sqrt{\hat{E}_1\hat{E}_2}}
%}, \nonumber \\ & & \hat{E}_i \equiv \sqrt{m_i^2 + \vec{\hat{q}}^2}
%\end{eqnarray}

%\[
%m_q = 90 \mbox{MeV}\, m_s = 260 \mbox{MeV}\, M_\pi = 140 \mbox{MeV}\, M_K
%= 490 \mbox{MeV}\;
%\]

%\begin{tabular}{|r|c|c|c|c|c|c|}
%\hline
%$\beta$ MeV & $f_\pi$ & $f_K$ & $<qq>^{(1/3)}$ & $<r^2_\pi>$ &
%$<r^2_{K^+}>$ & $<r^2_{K^0}>$ \\
%\hline
%.24  &  88.2 &  91.9 &  -158.  &    0.60 &     0.39  &      -0.066  \\
%
%.26  &  92.8 &  97.7  & -168.  &    0.52  &    0.34  &      -0.053  \\
%
%.28  &  97.1 & 103.2  & -177. &     0.45  &    0.29   &     -0.044   \\
%
%.30  & 101.3 & 108.3 &  -186.  &    0.41  &    0.26    &    -0.037  \\
%
%.32  & 105.3 & 113.3 &  -195.  &    0.36 &     0.23   &     -0.031   \\
%\hline
%exp. & 93.2 &  113. &   -250.  &    0.43 &     0.34   &
%-0.054 \\      &  MeV    & $\pm$ 2. MeV  &  $\pm$ 25. MeV &   $\pm$ 0.02
%fm$^2$ &$\pm$ 0.05 fm$^2$ & $\pm$0.026  fm$^2$\\
%\hline
%\end{tabular}


\begin{references}
\bibitem[1]{quarkmodels} A.~De R\'{u}jula, H.~Georgi, and S.~L.~Glashow,
Phys. Rev. D {\bf 12}, 147 (1975); O.~W.~Greenberg, Ann. Rev. of Nucl. Part.
Phys. {\bf 28}, 327 (1978); D.~P.~Stanley and D.~Robson, Phys. Rev. D {\bf
21}, 3180 (1980); S.~Godfrey and N.~Isgur, Phys. Rev. D {\bf 32}, 189
(1985); S.~Capstick and N.~Isgur, Phys. Rev. D {\bf 34}, 2809 (1986).

\bibitem[2]{Dzi} Z.~Dziembowski and L.~Mankiewicz, Phys. Rev. Lett.
{\bf 55}, 2175 (1987); Z.~Dziembowski, Phys. Rev. D {\bf 37}, 2030 (1988); 
{\bf 37}, 768 (1988); {\bf 37}, 778 (1988); Z.~Dziembowski, H.~J.~Weber,
L.~Mankiewicz, and A.~Szczepaniak, Phys. REv. D {\bf 39}, 3257 (1989);
C.-R. Ji and S.R.~Cotantch, Phys. Rev. D {\bf 41}, 2319 (1990);
C.-R. Ji, P.L.~Chung, and S.R.~Cotanch, Phys. Rev. D {\bf 45}, 4214 (1992).

\bibitem[3]{Coester} F.~Coester in {\it Proceedings of Nuclear and Particle
Physics on the Light Cone}, proceedings of the Workshop, Los Alamos, New
Mexico, 1988, edited by M.~Johnson and L.~Kisslinger, (World Scientific,
Singapore, 1989); P.~L.~Chung, F.~Coester, and W.~N.~Polyzu, Phys. Lett. B
 {\bf 205}, 545 (1988); P.~L.~Chung and F.~Coester, Phys. Rev. D {\bf 44},
 229 (1991), P.~L.Chung (unpublished).



\bibitem[4]{Shifman} A.~J.~C.~Hey and J.~E.~Mandula, Nucl. Phys. B {\bf
198}, 237 (1982); J.~R.Finger, J.~E.~Mandula, {\it ibid.} B {\bf 199},
168 (1982); A. Le Yaouanc, {\it et al.},
 Phys. Rev. D {\bf 31}, 137 (1985);
 E.~Shuryak, Nucl. Phys. B {\bf 328}, 85 (1989);
A.~G.~Williams, G.~Krein, and C.~D.~Roberts, Ann. Phys. {\bf 210}, 464
(1991).

\bibitem[5]{Bogolubow} N.~N.~Bogoliubov, Ann. Inst. Henri Poincar\'{e},
{\bf 8}, 163 (1963); A. Le Yaounac {\it et al.}, Phys. Rev. D {\bf 9},
2636 (1974); C.~Hayne and N.~Isgur, Phys. Rev. D {\bf 25}, 1944 (1982);
Z.~P.~Li and F.~E.~Close, Phys. Rev. D {\bf 43}, 2161 (1991).

\bibitem[6]{van} R.~Van Royen and V.~F.~Weisskopf, Nuov. Cim. {\bf 50},
617 (1967); {\bf 51}, 583 (1967); C.~H.~Llewellyn-Smith, Ann. Phys. {\bf
53}, 521 (1969); R.~P.~Feynman, M.~Kislinger, and F.~Ravndal, Phys. Rev.
{\bf 3}, 2706 (1971).

\bibitem[7]{Dirac} P.~A.~M.~Dirac, Rev. Mod. Phys. {\bf 21}, 392 (1949);
 {\it Lectures on Quantum Mechanics} (Belfer Graduate School of Science
 Monographs Series, Yeshiva University, New York, 1964).


\bibitem[8]{Smirnow} H.~Leutwyler and J.~Stern, Ann. Phys. {\bf 112}, 94
(1978); F.~Coester and
W.~N.~Polyzou, Phys. Rev. D {\bf 26}, 1348 (1982).

\bibitem[9]{others} L.D.~Faddeev, Teor. Mat. Fiz. {\bf 1}, 3 (1969);
I.~T.~Todorov,
in {\it Relativistic Action at a Distance : Classical and Quantum Aspects},
proceedings of the Barcelona Workshop, 1981, edited by J. Llosa (Lecture
Notes in Physics, Vol. 162) (Springer, Berlin, 1982); T.~Takabayasi, Prog.
Theor. Phys. {\bf 54}, 563 (1975);
, A.~Komar, Phys. Rev. D {\bf
18}, 1881 (1978); 1887 (1978); {\bf 18}, 3617 (1978);
S.~N.~Sokolov, Thor. Mat. Fiz. {\bf 36}, 193 (1978);
H.~Crater and P.~Van Alstine, Ann. Phys. {\bf 148}, 1308 (1983);
Phys. Rev. Lett. {\bf 53}, 1527 (1984); Phys. Rev. D {\bf 30},
2585 (1984).

\bibitem[10]{saz} H.~Sazdjian, Ann. Phys. {\bf 136}, 136 (1981);
Phys. Rev. D {\bf 33}, 3401 (1986).


\bibitem[11]{Winberg} S.~Weinberg, Phys. Rev. {\bf 150}, 1313 (1966);
J.~B.~Kogut and D.~E.~Soper, Phys. Rev. D {\bf 1}, 2901 (1970);
J.~D.~Bjorken, J.~B.~Kogut, and D.~E.~Soper, Phys. Rev. D {\bf 3}, 1382
(1971);
S.-J.~Chang, R.~G.~Root, and T.-M.~Yan, Phys. Rev. D {\bf 7}, 1133 (1973);
S.-J.~Chang and T.-M.~Yan {\it ibid.} 1147; T.-M.~Tan {\it ibid.} 1760;
T.-M.~Yan {\it ibid.} 1780; S.~J.~Brodsky, R.~Roskies, and R.~Suaya,
Phys. Rev. D {\bf 1}, 1035 (1970).

\bibitem[12]{DIS} G.~C.~Callan and D.~J.~Gross, Phys. Rev. Letters {\bf
22}, 156 (1969); D.~J.~Gross and C.~Llewellyn-Smith, Nucl. Phys. B {\bf 14},
 337 (1969), J.~D.~Bjorken and E.~A.~Paschos, Phys. Rev. D {\bf 1}, 3151
 (1970); D.~J.~Gross and S.~B.~Treiman, Phys. Rev. D {\bf 4}, 1059 (1971).


\bibitem[13]{LCFT} S.~J.~Brodsky,
in {\it Proceedings of the La Jolla Inst. Summer Workshop on QCD},
La Jolla (1978); S.~J.~Brodsky and G.~P.~Lepage in {\it Perturbative
Quantum Chromodynamics}, edited by A.~H.~Mueller, (World Scientific, 1989,
Singapore); C.~R.~Ji, in {\it Nuclear and Particle Physics on the Light
Cone}, edited by M.B.~Johnson and L.S.Kisslinger, (World Scientific, 1989,
Singapore).

\bibitem[14]{Karmanov}
 V.~A.~Karmanov and A.~V.~Smirmov, Nucl. Phys. A
{\bf 546}, 691 (1992).

\bibitem[15]{mock} N.~Isgur, Acta Phys. Pol. B {\bf 8}, 1081 (1977).

\bibitem[16]{PDG} Particle Data Group, Phys. Lett. B {\bf 239}, 1 (1990).

\bibitem[17]{Wigner} E.~Wigner, Ann. Math. {\bf 40}, 149 (1939);
P.~Moussa and R.~Stora, in {\it Methods in Subnuclear Physics},
proceedings of the 1968 Herceg-Novi International School on Elementary
Particle Reactions, Vol. 2 (New York, 1968); M.L.~Golberger and
K.~M.~Watson, {\it Collision Theory}, (Wiley, New York, 1964).

\bibitem[18]{ruscy} L.A.~Kondratyk and M.V.~Ternt'ev, Sov. J. Nucl. Phys.
{\bf 31}, 561 (1980).

\bibitem[19]{Deuteron} L.L.~Frankfurt and M.~I.Strikman, Nucl. Phys. B {\bf
148}, 107 (1979); I.~L.~Grach and L.~A.Kondratyuk, Sov. J. Nucl. Phys. {bf
39}, 198 (1984).

\bibitem[20]{ASAW} A.~Szczepaniak and A.~G.~Williams, Phys. Rev. D {\bf
47},  1175 (1993).

\bibitem[21]{SVZ} M.~A.~Shifman, A.~I.~Vainshtein, and V.~I.~Zakharov,
Nucl. Phys. B {\bf 147}, 385, 448, 519 (1979).

\bibitem[22]{last} In the Coester model of Ref.[3] for the light
quark we have used $m=330\mbox{MeV}$ and $\beta = 280\mbox{MeV}$,
values that give the best overall fit for the static properties of
$\pi$ and $K$.

\bibitem[23]{pionff} A.~Szczepaniak and A.G.~Williams, Phys. Lett. B {\bf
302}, 87 (1993); F.~Schlumpf, preprint SLAC-PUB-5998, Stanford, (1992);
A.V.~Radyushkin, preprint CABAF-TH-93-12.

\bibitem[24]{itzu} C.~Itzykson and J.-B.~Zuber, ``Quantum Field
Theory'', McGraw-Hill, New York, 1989.

\bibitem[25]{piondat} J.~Bebek {\it et al.} Phys. Rev. D {\bf 17}, 1693
(1978), S.~R.~Amendolia {\it et al.} Nucl. Phys. B {\bf 277}, 168 (1986).  

\bibitem[26]{kdata} S.~R.~Amendolia {\it et al.}, Phys. Lett. B {\bf 178},
435 (1986).


\end{references}
\end{document}